\documentclass[journal]{IEEEtran}

\usepackage{amssymb}
\usepackage{amsmath}
\usepackage{graphicx}
\usepackage{cite}
\usepackage{citesort}
\usepackage{balance}
\usepackage[utf8x]{inputenc}
\usepackage{amsmath,amssymb,amsthm}
\usepackage{xcolor}
\usepackage{graphicx,epstopdf}
\usepackage{epsfig}	
\usepackage[norule]{footmisc}

\usepackage{lipsum}

\newtheorem{theorem}{Theorem}

\newtheorem{lemma}{Lemma}

\newtheorem{corollary}{Corollary}

\makeatletter
\def\ScaleIfNeeded{%
\ifdim\Gin@nat@width>\linewidth \linewidth \else \Gin@nat@width
\fi } \makeatother

\begin{document}

\title{Wireless Energy Harvesting in a Cognitive Relay Network}

\author{Yuanwei~Liu,~\IEEEmembership{Student Member,~IEEE,}
        S.~Ali~Mousavifar,~\IEEEmembership{Student Member,~IEEE,}
        Yansha Deng,~\IEEEmembership{Student Member,~IEEE,}
        Cyril~Leung,~\IEEEmembership{Member,~IEEE,}
        Maged~Elkashlan,~\IEEEmembership{Member,~IEEE,}


\vspace{-0.2cm}

\thanks{The review of this paper was coordinated by Dr. Mehmet Can Vuran.}
\thanks{Y. Liu, Y. Deng, and M. Elkashlan are with the School of Electronic Engineering and Computer Science, Queen Mary University of London, London E1 4NS, UK (email: \{yuanwei.liu, yansha.deng, maged.elkashlan\}@qmul.ac.uk).}
\thanks{S. Ali Mousavifar and Cyril Leung are with the Department of Electrical and Computer Engineering, The University of British Columbia, Vancouver V6T 1Z4, Canada (email: \{seyedm, cleung\}@ece.ubc.ca).}
\thanks{This work was supported in part by the Natural Sciences and Engineering Research Council (NSERC) of Canada under Grant RGPIN 1731-2013.}}


\maketitle

\begin{abstract}
Wireless energy harvesting is regarded as a promising energy supply alternative for energy-constrained wireless networks. In this paper, a new wireless energy harvesting protocol is proposed for an underlay cognitive relay network with multiple primary user (PU) transceivers. In this protocol, the secondary nodes can harvest energy from the primary network (PN) while sharing the licensed spectrum of the PN. In order to assess the impact of different system parameters on the proposed network, we first derive an exact expression for the outage probability for the secondary network (SN) subject to three important power constraints: 1) the maximum transmit power at the secondary source (SS) and at the secondary relay (SR), 2) the peak interference power permitted at each PU receiver, and 3) the interference power from each PU transmitter to the SR and to the secondary destination (SD). To obtain practical design insights into the impact of different parameters on successful data transmission of the SN, we derive throughput expressions for both the delay-sensitive and the delay-tolerant transmission modes. We also derive asymptotic closed-form expressions for the outage probability and the delay-sensitive throughput and an asymptotic analytical expression for the delay-tolerant throughput as the number of PU transceivers goes to infinity. The results show that the outage probability improves when PU transmitters are located near SS and sufficiently far from SR and SD. Our results also show that when the number of PU transmitters is large, the detrimental effect of interference from PU transmitters outweighs the benefits of energy harvested from the PU transmitters.
\end{abstract}


\section{Introduction}\label{sec_intro}
Energy harvesting (EH), i.e., the process of extracting energy from the surrounding environment, has been proposed as an alternative method to supply energy and to prolong the lifetime of energy-constrained communication networks. A variety of harvestable energy sources such as heat, light, wave, and wind have been considered for EH in wireless networks \cite{sensor+energy_harvesing,vibration+energy_harvest,survey+energy_harvesting}.
Recently, harvesting energy from ambient radio frequency (RF) signals has received increasing attention due to its convenience in providing energy self-sufficiency to a low power communication system \cite{WPT+shinohara}. With recent advances in the technology of low power devices both in industry \cite{powercast} and academia \cite{parks2014turbocharging,harvesting_technology+Vullers2009684}, it is expected that harvesting energy from RF signals will provide a practically realizable solution for future applications, especially for networks with low power devices such as wireless sensor network (WSN) nodes \cite{cheng2013optimal,he2013energy}.

Wireless EH has been proposed in non-relay assisted as well as relay assisted networks \cite{EH_original+TS+PS,DTS+EH+CR,harvest1,harvest3,NAsir_harvest}. Two commonly used EH receiver architectures, namely power splitting (PS) receiver and time switching (TS) receiver, are proposed and studied in \cite{EH_original+TS+PS}. In a PS receiver, a fraction of the received signal power is used to harvest energy and the remainder is used to retrieve information. A TS receiver harvests energy from the received signal for a fraction of the time and retrieves the information from the received signal the rest of the time \cite{EH_original+TS+PS}. The fundamental tradeoff between harvesting energy and transmitting information in a non-relay assisted network over a variety of channel models is investigated in \cite{DTS+EH+CR,harvest1}. Two practical receivers based on PS are proposed and the rate of information transfer and the harvested energy are studied in \cite{harvest3}. Inspired by the TS and PS receiver architectures, two EH relaying protocols, namely time switching relay (TSR) protocol and power splitting relay (PSR) protocol, are proposed for an amplify-and-forward dual-hop network in \cite{NAsir_harvest}, where the energy constrained relay node harvests energy from RF signals of the source and uses the harvested energy to forward the information from the source to the destination. The delay-tolerant and delay-limited throughputs for EH relaying protocols are analyzed in \cite{NAsir_harvest}.

Cognitive radio (CR) is a promising technology which aims to achieve better spectrum utilization \cite{CR+Mitola}. Since point-to-point communications in CR networks is well established in the existing literature, recent research on CR mainly focused on cooperative relaying. In \cite{lee2006cth17}, a scenario with a single source-destination pair, assisted by a group of cognitive relay nodes, is considered. In \cite{lee2011outage}, considering a relay selection criterion, the outage probability of cognitive relay networks is evaluated. It is also shown in \cite{lee2011outage} that the diversity order of selection in cognitive relay networks is the same as in conventional relay networks.  Note that the aforementioned works mainly consider single primary transmitter and receiver. For the multiple primary transmitters and receivers case, the outage performance of cognitive relay networks with single antenna and multiple-input multiple-output (MIMO) is investigated in \cite{Trung+maged+multiple} and \cite{yeoh2013transmit}, respectively.

Recently, based on the advantages of the aforementioned two concepts, energy harvesting has been introduced to CR networks in \cite{EH+Overlay,underlay+overlay+EH,EE+Overlay3,EH+Overlay2,harvest+CR_Lu,harvest_CR_Underlay_XU_MIMO,harvest+CR_wang,harvest_CR_underlay}. In \cite{EH+Overlay}, EH and opportunistic spectrum access (OSA) are jointly studied, where OSA refers to a paradigm in which secondary users (SUs) utilize unused PU spectrum for transmissions. The throughput in a non-relay assisted CR with EH and overlay spectrum access is studied in \cite{underlay+overlay+EH}. An optimal spectrum sensing policy which aims to maximize the throughput subject to an energy constraint and a collision constraint is investigated in \cite{EE+Overlay3}. In \cite{EH+Overlay2}, an optimal mode selection policy is proposed to maximize total throughput in the cognitive radio network (CRN), where the SU switches between EH mode and OSA mode. In \cite{harvest+CR_Lu}, the throughput in an amplify-and-forward (AF) cognitive relay network is maximized subject to transmission time and energy constraints at the CRN. In \cite{harvest_CR_Underlay_XU_MIMO}, the sum of harvested powers at multiple energy harvesting SU receivers in a MIMO underlay CR is maximized, subject to satisfying two constraints (i.e., a target minimum-square-error at multiple CR information receivers and the peak interference power constraint at the primary network receivers). In \cite{harvest+CR_wang}, the outage probability is analyzed for a non-relay assisted secondary network (SN) which shares the spectrum and harvests energy while assisting a primary transmission. The results in \cite{harvest+CR_wang} show that the SN can harvest sufficient energy from RF signals to relay the information for the primary network (PN) as well as transmit its own information. In \cite{harvest_CR_underlay}, an EH protocol for a non-relay assisted SN in underlay CR with a single SU transmitter and multiple energy-constrained SU receivers is studied.

\subsection{Motivation and Contributions}
The motivation behind adopting wireless energy harvesting in underlay spectrum sharing networks with multiple PU transceivers can be described as follows: 1) From the perspective of SU, the RF signal from the PUs can be regarded as a constant energy source due to the concurrent transmission between the PN and the SN; 2) The transmit power at SU must remain below a predetermined threshold due to the interference power constraint from PN. This well matches the notion of wireless energy harvesting as a promising technique for low-power devices due to the limitation of power transfer efficiency; and 3) It is realistic to consider multiple PU transceivers to operate simultaneously in a large scale CR network. Increasing the number of PU transmitters increase the energy harvested by the SU. However, as the number of PU transmitters increases, the interference from PU transmitters on the SU also increases. Therefore, it is important to study the tradeoff between the benefits of energy harvesting and the harmful effects of interference.

In this paper, we propose a wireless EH protocol for a decode-and-forward (DF) cognitive relay network with multiple PU transceivers. With this protocol, both the energy constrained SS and secondary relay (SR) nodes can harvest energy from the RF signals of multiple PU transmitters to support information transmission. The impact of EH and PU interference on the outage probability and the throughput in the SN are studied.

The main contributions of this paper are summarized as follows:
\begin{itemize}

 \item We derive an exact expression for the system outage probability in the cognitive relay network subject to three power constraints: 1) the maximum transmit power at SS and SR based on the harvested energy; 2) the peak interference power at each PU receiver; and 3) the interference power from multiple PU transmitters at SR and SD.
 \item We derive analytical expressions for the throughput both in the delay-sensitive and the delay-tolerant transmission modes.
 \item We derive asymptotic closed-form expressions for the outage probability and delay-sensitive throughput and an asymptotic expression for the delay-tolerant throughput as the number of PU transmitters and receivers goes to infinity.

  \item We show that there exists an optimal number of PU transceivers that achieves the minimum outage probability and the maximum throughput for the SN.
 \item We show that when the number of PU transmitters is large, the negative impact caused by the interference from the PU transmitters outweighs the positive impact brought by the EH from the PU transmitters at the SN.
 \end{itemize}
\subsection{Notation and Organization}

The remainder of the paper is organized as follows. In Section \ref{sec_sys_model}, we present the system model and assumptions for the EH relaying protocol. In Section \ref{sec_analysis}, we derive exact expressions for the outage probability and the delay-tolerant and delay-sensitive throughputs. In Section \ref{sec_large_analysis}, we derive exact asymptotic expressions for the outage probability and the delay-tolerant and delay-sensitive throughputs as the number of PU transceivers goes to infinity. Illustrative results and conclusions are provided in Sections~\ref{sec_simualtion} and \ref{sec_concl}, respectively.

\vspace{-.2cm}\section{Network Model}\label{sec_sys_model}
\begin{figure}
\centering
\includegraphics[width=2.8in, height=2.05in]{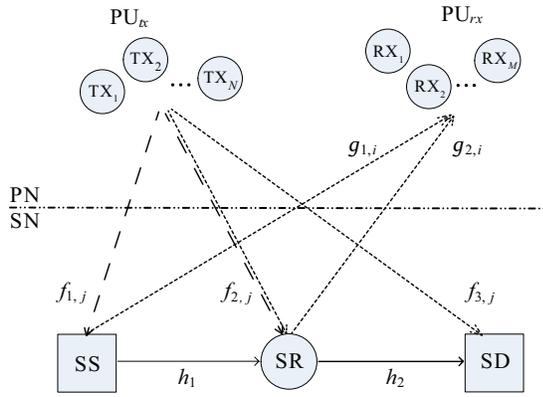}
 \caption{System model of the energy harvesting cognitive radio system. The EH links, interference links, and information links are illustrated with the dashed, dotted, and solid lines, respectively.}\label{fig_system_model}
\end{figure}

As shown in Fig. \ref{fig_system_model}, we consider an underlay cognitive relay network where an energy constrained SS transmits to an energy sufficient SD through an energy constrained SR using the licensed PU spectrum. The primary network consists of $M$ primary transmitters ($\text{PU}_{tx}$) and $N$  primary receivers ($\text{PU}_{rx}$)~\footnote{We assume that PU receivers apply multi-user detectors to cancel interference from different PU transmitters \cite{multi_user_interference}.}. We assume that all PU transmitters are closely located in one center point and all PU receivers are closely located in another center point \cite{Trung+maged+multiple,yeoh2013transmit,cluster+1}. We assume that there is no direct link between SS and SD due to deep fading \cite{bletsas2006simple,krikidis2012full} and the communication between SS and SD can only be completed with the help of SR \cite{NAsir_harvest}. Both of the SS and SR only use the energy harvested from RF signals of PU transmitters. We consider a communication system with rechargeable storage ability at SS and SR. All the energy harvested during the energy harvesting time slot is used for information transmission \cite{NAsir_harvest,Batteryless4}. From the implementation point of view, this rechargeable storage unit can be a supercapacitor or a short-term/high-efficiency battery to support the switching between energy harvesting and information transmission \cite{BatteryLess+MIMO}. Note that the SN shares the spectrum with the PN in an underlay paradigm, which means that secondary users can perform concurrent transmission as long as the interference at PU does not exceed a peak permissible threshold, denoted by $P_{\mathcal{I}}$. Since there is no sensing in an underlay paradigm, we assume that the energy required to receive/process information is negligible compared to the energy required for information transmission \cite{NAsir_harvest,Goldstone+Energy+constrainedmodulation,harvest+wireless}. We consider interference-limited case where the interference power caused by PU transmitters at SR and SD is dominant relative to the noise power \cite{Trung+maged+multiple}. All channels are assumed to be quasi-static Rayleigh fading channels where the channel coefficients are constant for each transmission block but vary independently between different blocks.

As shown in Fig. \ref{fig_system_model}, we denote the channel gain coefficients from SS and SR to the $i$th $\text{PU}_{rx}$ by
$g_{1,i}$ and $g_{2,i}$ for $i=1,2,\dots,M$, respectively. We denote the channel gain coefficients from SS to SR and from SR to SD by $h_{1}$ and $h_{2}$, respectively. And the channel gain coefficients from the $j$th $\text{PU}_{tx}$ to SS, SR, and SD, are denoted by $f_{1,j}$, $f_{2,j}$, and $f_{3,j}$ for $j=1,2,,\dots,N$, respectively. We denote the distance between the $j$th $\text{PU}_{tx}$ and SS, SR, and SD as $d_{1,j}$, $d_{2,j}$, and $d_{3,j}$, respectively. The distance between SS and SR and the $i$th $\text{PU}_{rx}$ are denoted as $d_{4,i}$ and $d_{5,i}$, respectively. And the distances between SS and SR and between SR to SD are denoted as $d_{6}$ and $d_{7}$, respectively. The link gain realizations $|h_{1}|^2$, $|h_{2}|^2$, $|g_{1,i}|^2$, $|g_{2,i}|^2$, $|f_{1,j}|^2$, $|f_{2,j}|^2$, and $|f_{3,j}|^2$ are exponentially distributed with parameters $\lambda_{1}$, $\lambda_{2}$, $\omega_{1,i}$, $\omega_{2,i}$, $\nu_{1,j}$, $\nu_{2,j}$, and $\nu_{3,j}$, respectively; for example, $\lambda_{1}=d_{6}^{-m}$, where $m$ is the path loss factor. The link gains from the PU transmitters to the SS are assumed to be identically distributed as well as those to the SR and SD \cite{Trung+maged+multiple}, i.e., $\nu_{1,j}=\nu_{1}$, $\nu_{2,j}=\nu_{2}$, and $\nu_{3,j}=\nu_{3}$ for $j=1,2,\dots,N$. Similarly, the link gains from SS to the PU receivers are assumed to be identically distributed as well as those from SR to PU receivers, i.e., $\omega_{1,i}=\omega_{1}$, $\omega_{2,i}=\omega_{2}$ for $i=1,2,\dots,M$. \footnote{To keep the analysis tractable, shadow fading is not considered in this paper.}

\begin{figure} [!h]
\centering
\includegraphics[width=3in, height=1.3in]{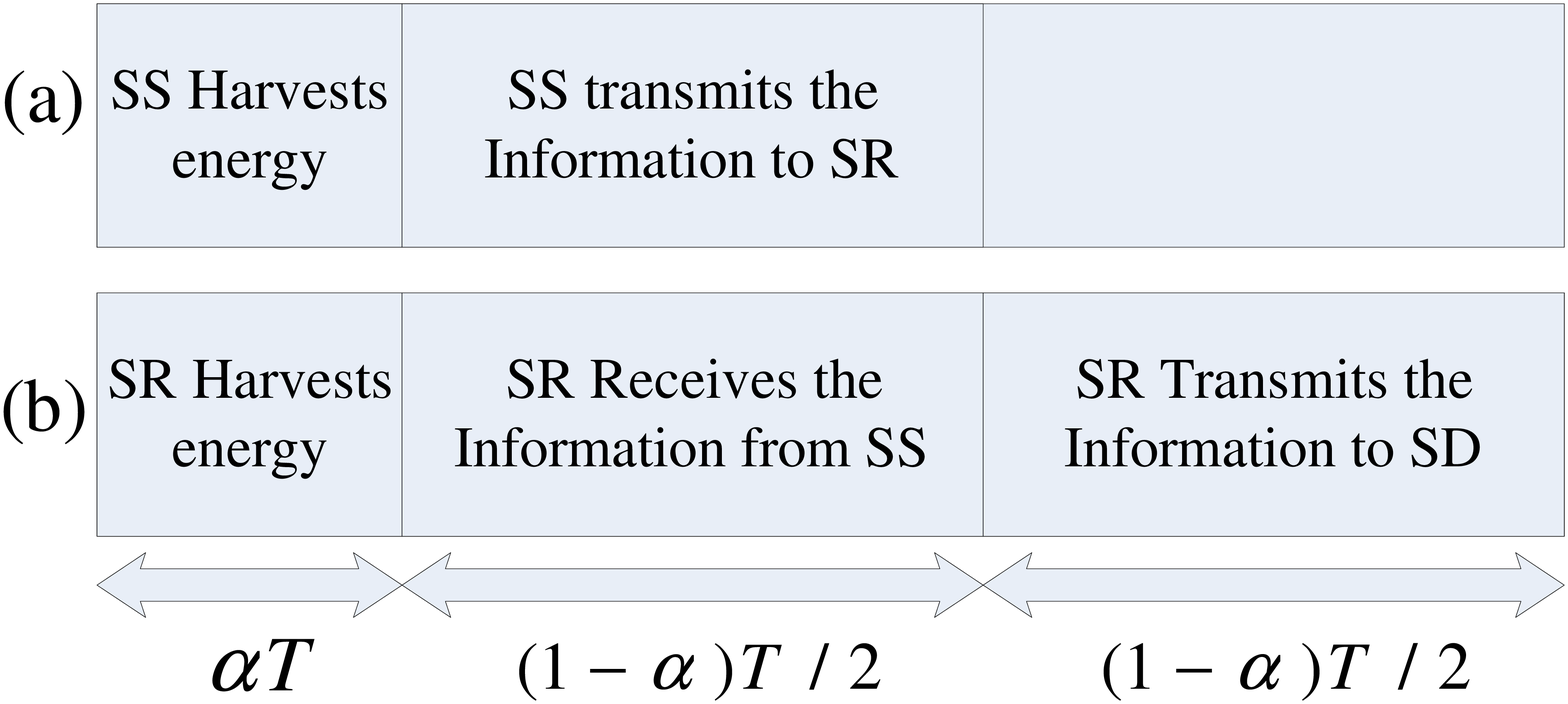}
 \caption{(a) illustrates the protocol at SS in one EH-IT time slot and (b) illustrates the protocol at SR in one EH-IT time slot.}\label{fig_protocol_S1}
\end{figure}

As shown in Fig. \ref{fig_protocol_S1}, the SS and SR harvest energy from RF signals of PU transmitters for a duration of $\alpha T$ at the beginning of each energy harvesting-information transmission (EH-IT) time slot, where $T$ is the duration of one EH-IT time slot and $0<\alpha<1$. We assume that regardless of how much energy is harvested at SS and SR in the EH time slot, they can store it in a storage device (e.g., a supercapacitor or a short-term/high-efficiency battery). Subsequent to the harvesting period, SS transmits information to SR for a duration $\beta(1-\alpha)T$, where $0<\beta<1$. Then, SR forwards the information to SD for a duration of $(1-\beta)(1-\alpha)T$. For simplicity, we assume $\beta=\frac{1}{2}$ in this network.

The energy harvested using the TS receiver architecture at SS and SR are given by \footnote{In this work, we assume that the power that can be transferred to SS and SR is greater than the activation threshold and the energy harvesting circuits at SS and SR are always activated. }
\begin{equation}
E_{h_{s}}=\eta P_{PU_{tx}} \sum_{j=1}^{N}|f_{1,j}|^2\alpha T,
\label{eq_Ehs_s1}
\end{equation}
\noindent and
\begin{equation}
E_{h_{r}}=\eta P_{PU_{tx}} \sum_{j=1}^{N}|f_{2,j}|^2\alpha T,
\label{eq_Ehr_s1}
\end{equation}

\noindent respectively, where $0<\eta< 1$ is the energy conversion efficiency \cite{harvest+wireless}, and $P_{PU_{tx}}$ is the transmit power of all PU transmitters). We consider constant transmit power of SS and SR during the IT time slot, therefore, the maximum powers that SS and SR can transmit at based on the harvested energy are $\frac{E_{h_{s}}}{(1-\alpha)T/2}$ and $\frac{E_{h_{r}}}{(1-\alpha)T/2}$, respectively. Therefore, the transmit power at SS and SR are given by
\begin{equation}
P_{s}=\min(\frac{E_{h_{s}}}{(1-\alpha)T/2},\frac{P_{\mathcal{I}}}{\max|g_{1,i}|^2}),
\label{eq_Ps_s1}
\end{equation}
\noindent and
\begin{equation}
P_{r}=\min(\frac{E_{h_{r}}}{(1-\alpha)T/2}, \frac{P_{\mathcal{I}}}{\max|g_{2,i}|^2}),
\label{eq_Pr_s1}
\end{equation}
\noindent respectively.

As the SS and SR share the same spectrum with the PUs, the transmitted signals from SS and SR cause interference to each $\text{PU}_{rx}$. Constraints on the transmit power of SS and SR are imposed in order that the interference power to the PU does not exceed $P_{\mathcal{I}}$.

Then signal to interference ratio (SIR) at SR and SD are obtained as
\begin{align}
\Gamma_{R}= \min (\rho {Z_1},\frac{{{P_{\cal I}}}}{{{Y_1}}})\frac{{{X_1}}}{{{Z_2}}},
\label{eq_gamma_R}
\end{align}
\noindent and
\begin{align}
\Gamma_{D}= \min (\rho {Z_2},\frac{{{P_{\cal I}}}}{{{Y_2}}})\frac{{{X_2}}}{{{Z_3}}},
\label{eq_gamma_D}
\end{align}
\noindent respectively, where $\rho=\frac{2 \eta \alpha}{(1-\alpha)}$, $X_{1}=|h_{1}|^2$, $X_{2}=|h_{2}|^2$,
$Z_{1}=\sum_{j=1}^{N}P_{PU_{tx}}|f_{1,j}|^{2}$, $Z_{2}=\sum_{j=1}^{N}P_{PU_{tx}}|f_{2,j}|^{2}$, $Z_{3}=\sum_{j=1}^{N}P_{PU_{tx}}|f_{3,j}|^{2}$, ${Y_1} = \mathop {\max }\limits_{i = 1,...,M} |{g_{1,k}}{|^2}$, and
${Y_2} = \mathop {\max }\limits_{i = 1,...,M} |{g_{2,k}}{|^2}$. Note that $\Gamma_{R}$ and $\Gamma_{D}$ are dependent random variables because they both are functions of $Z_{2}$.
\section{Exact Performance Analysis}\label{sec_analysis}
In this section, we derive expressions for the outage probability and throughput for the secondary network with the proposed energy harvesting protocol. These expressions provide practical design insights into the impact of various parameters on the performance of the secondary network.
\subsection{Outage Probability}
The outage probability $P_{out}$, is defined as the probability that the equivalent SIR at each hop is below a threshold value, $\gamma_{th}$. In our paper, the DF relay assisted spectrum sharing network is considered to be in outage if only one of the links is suffering from an outage, i.e.,
\begin{align}
P_{out} (\gamma_{th}) =1- \Pr\{\Gamma_{R} \geq \gamma_{th}, \Gamma_{D}\geq\gamma_{th}\},
\label{eq_Pout}
\end{align}
\noindent where $\Gamma_{R}$ and $\Gamma_{D}$ denote the SIR RVs at SR and SD, and are given in \eqref{eq_gamma_R} and \eqref{eq_gamma_D}, respectively.

To facilitate the outage probability, we first derive the cumulative distribution function (CDF) and the probability distribution function (PDF) of ${Z_p }$, $ {p  \in \left\{ {1,2,3} \right\}}$ and ${Y_q }$, $q  \in \left\{ {1,2} \right\}$.
Note that each ${Z_p }$ is the sum of $N$ independent exponential RVs and its distribution is chi-square distribution. The PDF and CDF of ${Z_p }$  are expressed as
\begin{align}
{f_{{Z_p }}}({z_p }) = \frac{{z_p ^{N - 1}{e^{ - \frac{{{z_p }}}{{{P_{P{U_{tx}}}}{\nu _p }}}}}}}{{\Gamma (N){{({P_{P{U_{tx}}}}{\nu _p })}^N}}},
\label{eq_chi_pdf}
\end{align}
\noindent and
\begin{align}
{F_{{Z_p }}}({z_p }) =\frac{{\Gamma (N,\frac{{{z_p }}}{{{P_{P{U_{tx}}}}{\nu _p }}})}}{{\Gamma (N)}},\label{eq_chi_cdf}
\end{align}
\noindent respectively, where $\Gamma(.)$ and $\Gamma(.,\hspace{-.05cm}.)$ denote the gamma and incomplete gamma functions, respectively.
The PDF and CDF of ${Y_q }$ are expressed as
\begin{align}
{f_{{Y_q }}}({y_q }) = \frac{M}{{{\omega _q }}}\sum\limits_k^{M - 1} {M-1 \choose k} {( - 1)^k}{e^{ - (\frac{{k + 1}}{{{\omega _q }}}){y_q }}},
\label{eq_max_pdf}
\end{align}
\noindent and
\begin{align}
{F_{{Y_q }}}({y_q }) = {(1 - {e^{ - \frac{{{y_q }}}{{{\omega _q }}}}})^M},
\label{eq_max_cdf}
\end{align}
\noindent respectively. We assume that the distances between the PU transmitters are relatively smaller than the distances between the PU transmitters and SS, SR, and SD. The same assumption is used for PU receivers with respect to SS, SR, and SD.

\noindent

The exact expression for the outage probability is given in the following theorem.

\begin{theorem}
The outage probability of the SN with the proposed energy harvesting protocol is given by
\begin{align}
P_{out}(\gamma_{th}) =& 1- \int_{0}^{\infty}(\mathcal{J}_{R,I}+\mathcal{J}_{R,II})\nonumber\\
&\times(\mathcal{J}_{D,I}+\mathcal{J}_{D,II})\frac{z_{2}^{N-1}e^{-\frac{z_{2}}{P_{PU_{tx}}\nu_{2}}}}{\Gamma(N)(P_{PU_{tx}}\nu_{2})^N}dz_{2},
\label{eq_Pout_simplfied}
\end{align}
\noindent where $\mathcal{J}_{R,I}$, $\mathcal{J}_{R,II}$, $\mathcal{J}_{D,I}$, and $\mathcal{J}_{D,II}$ are defined in Appendix~\ref{sec_appendixA}.
\end{theorem}
\begin{IEEEproof}
See Appendix \ref{sec_appendixA}.
\end{IEEEproof}

\subsection{Throughput}\label{sec_throughput}
In this section, we evaluate the throughput in two transmission modes: delay-sensitive mode and delay-tolerant mode. The evaluation of throughput provides insight into practical implementations and challenges of EH cognitive relay network.

\subsubsection{Delay-sensitive Transmission}
In this mode, SS transmits information to SR at a fixed rate, denoted by $R_{ds}$, where $R_{ds}\triangleq \text{log}_{2}(1+\gamma_{th})$. The throughput in delay-sensitive mode is expressed as \cite{NAsir_harvest}
\begin{eqnarray}
\tau_{ds} &=& \frac{\frac{(1-\alpha)T}{2}}{T}R_{ds}(1-P_{out}(\gamma_{th}))\nonumber \\
&=&\frac{(1-\alpha)}{2}R_{ds}(1-P_{out}(\gamma_{th})),
\label{eq_Throughput_DL}
\end{eqnarray}
\noindent where $\tau_{ds}$ is equal to the rate of the successful transmissions during the transmission time, $\frac{(1-\alpha)T}{2}$, when SIR requirement ($\gamma_{th}$) is satisfied at SR and SD. Note, the system outage probability is obtained from \eqref{eq_Pout_simplfied}.
\subsubsection{Delay-Tolerant Transmission}
In this mode, users can transmit messages reliably while the transmission rate is less than or equal to the ergodic capacity $C_{erg}$, which is defined as \cite{Proakis+capacity}
\begin{equation}
C_{erg}= \mathbb{E}\{\text{log}_{2}(1+\Gamma_{th})\}, \nonumber \\
\label{eq_Ergodic_cap_original}
\end{equation}
\noindent where $\mathbb{E}\{.\}$ denotes the expected value of an argument and $\Gamma_{th}$ is a random variable defined as $\gamma_{th}= \text{min}(\gamma_{R}, \gamma_{D})$. We obtain delay-tolerant throughput, denoted by $\tau_{dt}$, as
\begin{eqnarray}
\tau_{dt} = \frac{\frac{(1-\alpha)T}{2}}{T}C_{erg}= \frac{(1-\alpha)}{2\text{ln}2}\int_{0}^{\infty}\frac{1-F_{\Gamma_{th}}(x)}{(1+x)}dx,
\label{eq_Throughput_DT}
\end{eqnarray}
\noindent where $F_{\Gamma_{th}}(\gamma_{th})$ denotes the CDF of $\gamma_{th}$ and it can be evaluated by \eqref{eq_Pout_simplfied}, i.e., $F_{\Gamma_{th}}(\gamma_{th})= P_{out}(\gamma_{th})$. In contrast to delay-sensitive mode, where SS is expected to transmit at a fixed rate to satisfy a certain outage probability, in delay-tolerant transmission mode, SS can transmit at any rate equal or less than the evaluated ergodic capacity.

\section{Large System Analysis}\label{sec_large_analysis}
In this section, we analyze the outage probability, the delay-sensitive throughput, and the delay-tolerant throughput as the number of PU transceivers grows to infinity. As $N$ increases, more interference imposed on SR and SD may cause outages. On the other hand, as $N$ increases, SS and SR can potentially harvest more energy from multiple PU transmitters. In addition, as $M$ increases, SS and SR can impose interference at more number of the PU receivers. The tradeoff between the positive impact brought by the energy harvesting from multiple PU transmitters and the negative influence brought by the interference of the multiple PU transmitters on SR and SD can result in optimal values for $M$ and $N$ which correspond to the minimum outage probability and the maximum throughput.

To do so, we will first look at the distribution of $Z_{1}$, $Z_{2}$, and $Z_{3}$ as $N\rightarrow\infty$. Since $Z_1$ is independent and non-identically distributed exponential RVs, the distribution of $Z_1$ is asymptotically normal as $Z_1\to\infty$. Using law of large numbers, we have
\begin{align}
    Z_1\mathop \to \limits^{d}N P_{PU_{tx}}{\nu _1},
    \label{MassiveZ 1}
\end{align}
\noindent where $A\mathop \to \limits^{d}B$ denotes convergence in distribution of a rv A to a rv B. Similarly, we have
\begin{align}\label{MassiveZ 2}
    Z_2 \mathop \to \limits^{d}N P_{PU_{tx}}{\nu _2}
\end{align}
\noindent and
\begin{align}\label{MassiveZ 3}
    Z_3\mathop \to \limits^{d} N P_{PU_{tx}} {\nu _3}.
\end{align}
In the same manner, we obtain the distribution of $Y_{1}$ and $Y_{2}$ as $M\rightarrow\infty$. Since $Y_1$ is the maximum of $M$ independent and non-identically distributed exponential RVs, the distribution of $Y_1$ is asymptotically normal, as $Y_1\to\infty$. From \cite[Proposition~1]{hesami2011limiting}, we have
\begin{align}\label{MassiveY 1}
    Y_1 \mathop \to \limits^{d} {\omega _1} + {\omega _1}\ln M + {{\overline{Y}_1}} ,
\end{align}
where $\overline{Y}_{1}$ is the normal distribution with $\mathcal{N}(0, 2{\omega _1}^2
)$.
Similarly, we obtain
\begin{align}\label{MassiveY 2}
    Y_2
    \mathop \to \limits^{d}
   {\omega _2} + {\omega _2}\ln M +  {{\overline{Y}_2}} ,
\end{align}
where $\overline{Y}_{2}$ is the normal distribution with $\mathcal{N}(0, 2{\omega _2}^2)$. Having the distribution of $Y_{1}$, $Y_{2}$, $Z_{1}$, $Z_{2}$, and $Z_{3}$ for large number of PU transceivers, we rewrite SIR in \eqref{eq_gamma_R} and \eqref{eq_gamma_D} at SR and at SD as
\begin{align}\label{R_SIR}
    {\Gamma _{R}^{\infty}}
    \mathop \sim \limits^{d}
   \min ({\rho}N P_{PU_{tx}} {\nu _1},\frac{{{P_I}}}{{{\omega _1} + {\omega _1}\ln M +  {{\overline{Y}_1}} }})\frac{{{X_1}}}{{{P_{P{U_{tx}}}}N{\nu _2}}}
\end{align}
\noindent and
\begin{align}\label{D_SIR}
    {\Gamma _{D}^{\infty}}
    \mathop \sim \limits^{d}
   \min ({\rho}N P_{PU_{tx}} {\nu _2},\frac{{{P_I}}}{{{\omega _2} + {\omega _2}\ln M +  {{\overline{Y}_2}} }})\frac{{{X_2}}}{{{P_{P{U_{tx}}}}N{\nu _3}}},
\end{align}
\noindent respectively. We use the SIR in \eqref{R_SIR} and \eqref{D_SIR} to obtain the outage probability for large number of PU tranceivers in the next section.
\subsection{Outage probability}
The asymptotic outage probability of the SN is given by
\begin{align}\label{P_out_large_1}
P_{out}^{\infty}(\gamma_{th}) = 1- \Pr\{\Gamma_{R}^{\infty}\geq \gamma_{th}, \Gamma_{D}^{\infty}\geq\gamma_{th}\}.
\end{align}
In contrast to \eqref{eq_Pout}, $\Gamma_{R}^{\infty} $ and $\Gamma_{D}^{\infty}$ in \eqref{P_out_large_1} are independent and therefore, we can write the second term in \eqref{P_out_large_1} as the product of two terms.
\begin{theorem}
As the number of transceivers goes to infinity, the outage probability of the proposed cognitive relay network is given by in closed form
\begin{eqnarray}\label{P_out_large_2}
{P_{out}^{\infty}}(\gamma_{th})= 1 -\Theta _R \Theta _D,
\end{eqnarray}

\noindent where ${{\Theta _R}}=\Pr \{ {\Gamma _{R}^{\infty}} \ge {\gamma _{th}}\} $ and ${{\Theta _D}}=\Pr \{ {\Gamma _{D}^{\infty}} \ge {\gamma _{th}}\} $, respectively.
\end{theorem}
\begin{IEEEproof}
The term ${{\Theta _R}}$ is obtained in \eqref{Gamma_R_large_2} at the top of next page. The terms in \eqref{Gamma_R_large_2_l3} can be obtained using the CDF and PDF expressions for $X_{1}$ and ${\overline{Y}_{1}}$.
\begin{figure*}[!t]
\begin{align}
\Theta _R = &\Pr \{ {\Gamma _R^{\infty}} \ge {\gamma _{th}}\}\nonumber\\
=&\Pr \left\{ {{X_1} \ge \frac{{{P_{P{U_{tx}}}}N{\nu _2}{\gamma _{th}}\left( {{\omega _1} + {\omega _1}\ln M + {\overline{Y}_1}} \right)}}{{{P_I}}},{\overline{Y}_1} \ge \frac{{{P_I}}}{{\rho N{P_{P{U_{tx}}}}{\nu _1}}} - {\omega _1} - {\omega _1}\ln M} \right\}\nonumber\\
& + \Pr \left\{ {{X_1} \ge \frac{{{\nu _2}{\gamma _{th}}}}{{\rho {\nu _1}}}} \right\}\Pr \left\{ { - {\omega _1} - {\omega _1}\ln M \le {\overline{Y}_1} \le \frac{{{P_I}}}{{\rho N{P_{P{U_{tx}}}}{\nu _1}}} - {\omega _1} - {\omega _1}\ln M} \right\}\label{Gamma_R_large_2_l2}\\
=&\int_{u_{R}}^\infty  {\left( {1 - {F_{{X_1}}}\left( {\frac{{\left( {{\omega _1} + {\omega _1}\ln M + {{\overline{Y}_1}} } \right){P_{P{U_{tx}}}}N{\nu _2}{\gamma _{th}}}}{{{P_I}}}} \right)} \right)} f\left( {{{\overline{y}}_1}} \right)d{{\overline{y}}_1}\nonumber\\
&+ {e^{ - \frac{{{\nu _2}{\gamma _{th}}}}{{{\lambda _1}\rho {\nu _1}}}}}\left( {\frac{1}{2}\left[ {\Phi \left( {\frac{{{u_R}}}{{2{\omega _1}}}} \right) - \Phi \left( {\frac{{{u_{{R^*}}}}}{{2{\omega _1}}}} \right)} \right]} \right)\label{Gamma_R_large_2_l3}\\
= &\frac{{{e^{ - \frac{{{P_{P{U_{tx}}}}N{\nu _2}{\gamma _{th}}{\omega _1}\left( {1 + \ln M} \right)}}{{{\lambda _1}{P_I}}}}}}}{{2{\omega _1}\sqrt \pi  }}\int _{{u_R}}^\infty {e^{ - \frac{{{\overline{Y}_1}^2}}{{4{\omega _1}^2}} - \frac{{{P_{P{U_{tx}}}}N{\nu _2}{\gamma _{th}}{\overline{Y}_1}}}{{{\lambda _1}{P_I}}}}}d{\overline{Y}_1}\nonumber\\
&+ \frac{1}{2}{e^{ - \frac{{{\nu _2}{\gamma _{th}}}}{{{\lambda _1}\rho {\nu _1}}}}}\left[ {\Phi \left( {\frac{{{u_R}}}{{2{\omega _1}}}} \right) - \Phi \left( {\frac{{{u_{{R^*}}}}}{{2{\omega _1}}}} \right)} \right]\label{Gamma_R_large_2}.
\end{align}
\hrulefill
\end{figure*}
Applying \cite[Eq 3.322.1]{TablOfIntegrals}, we can further simplify $\Theta _R$ obtained in \eqref{Gamma_R_large_2} as
\begin{align}\label{Gamma_R_large_3}
\Theta _R = &\frac{{{e^{{\omega _1}^2{\gamma _R}^2 - {\gamma _R}{\omega _1}\left( {1 + \ln M} \right)}}}}{2}\left( {1 - \Phi \left( {{\gamma _R}{\omega _1} + \frac{{{u_R}}}{{2{\omega _1}}}} \right)} \right)\nonumber\\
&+ \frac{1}{2}{e^{ - \frac{{{\nu _2}{\gamma _{th}}}}{{{\lambda _1}{\rho}{\nu _1}}}}}\left[ {\Phi \left( {\frac{{{u_R}}}{{2{\omega _1}}}} \right) - \Phi \left( {\frac{{{u_{{R^*}}}}}{{2{\omega _1}}}} \right)} \right],
\end{align}
where ${\gamma _R} = \frac{{{P_{P{U_{tx}}}}N{\nu _2}{\gamma _{th}}}}{{{\lambda _1}{P_I}}}$, $\Phi \left( x \right) = \frac{2}{{\sqrt \pi  }}\int_0^x {{e^{ - {t^2}}}} dt$, ${u_R} = \frac{{{P_I}}}{{\rho N{P_{P{U_{tx}}}}{\nu _1}}} - {\omega _1} - {\omega _1}\ln M$, and ${u_{R^*}} =  - {\omega _1} - {\omega _1}\ln M$. Similarly, we obtain ${{\Theta _D}}$ by substituting the parameters of \eqref{Gamma_R_large_3} with $ {{\nu _1}}\rightarrow {{\nu _2}}$,  ${{\omega _1}}\rightarrow {{\omega _2}}$, ${{\lambda _1}}\rightarrow {{\lambda _2}}$, and ${{\nu _2}}\rightarrow {{\nu _3}}$. Then we have
\begin{align}\label{Gamma_D_large_2}
\Theta _D =&\frac{{{e^{{\omega _2}^2{\gamma _D}^2 - {\gamma _D}{\omega _2}\left( {1 + \ln M} \right)}}}}{2}\left( {1 - \Phi \left( {{\gamma _D}{\omega _2} + \frac{{{u_D}}}{{2{\omega _2}}}} \right)} \right)\nonumber\\
&+ \frac{1}{2}{e^{ - \frac{{{\nu _3}{\gamma _{th}}}}{{{\lambda _2}\rho {\nu _2}}}}}\left[ {\Phi \left( {\frac{{{u_D}}}{{2{\omega _2}}}} \right) - \Phi \left( {\frac{{{u_{{D^*}}}}}{{2{\omega _2}}}} \right)} \right],
\end{align}
\noindent where ${u_D} = \frac{{{P_I}}}{{\rho {P_{P{U_{tx}}}}N{\nu _2}}} - {\omega _2} - {\omega _2}\ln M, {u_{D^*}} =  - {\omega _2} - {\omega _2}\ln M$, and ${\gamma _D} = \frac{{{P_{P{U_{tx}}}}N{\nu _3}{\gamma _{th}}}}{{{\lambda _2}{P_I}}}$. Substituting \eqref{Gamma_R_large_3} and \eqref{Gamma_D_large_2} into \eqref{P_out_large_2}, we obtain a closed-form expression for the outage probability.
\end{IEEEproof}
\subsection{Throughput}
We derive the delay-sensitive throughput and delay-tolerant throughput in this section. We use the expression derived in \eqref{P_out_large_2} to evaluate $\tau_{ds}^{\infty}$ and $\tau_{dt}^{\infty}$, respectively, as
\begin{equation}
\tau_{ds}^{\infty} =\frac{(1-\alpha)}{2}R_{ds}(1-P_{out}^{\infty}(\gamma_{th})),
\label{eq_Throughput_DL_large}
\end{equation}
 \noindent and
\begin{eqnarray}
\tau_{dt}^{\infty} = \frac{(1-\alpha)}{2\text{ln}2}\int_{0}^{\infty}\frac{1-F_{\Gamma_{th}}(x)}{(1+x)}dx,
\label{eq_Throughput_DT_Large}
\end{eqnarray}
\noindent where $F_{\Gamma_{th}}(\gamma_{th})$ can be evaluated using \eqref{P_out_large_2}, i.e., $F_{\Gamma_{th}}(\gamma_{th})= P_{out}^{\infty}(\gamma_{th})$.
\section{Numerical Results}\label{sec_simualtion}

\begin{figure} [t!]
\begin{center}
\includegraphics[width=3.5in, height=2.5in]{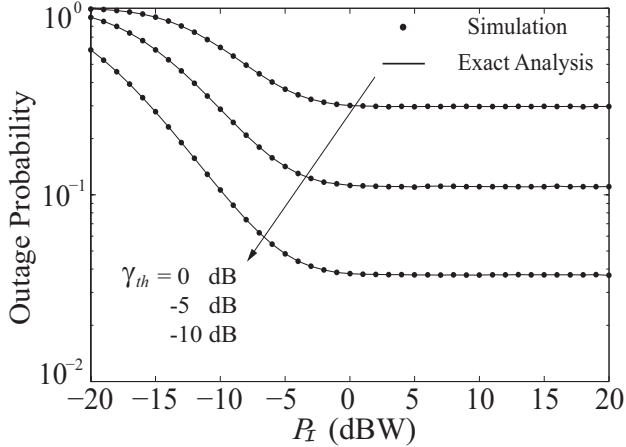}
 \caption{Outage probability as a function of $P_{\mathcal{I}}$, with $M=3$, $P_{PU_{tx}}=0$~dBW.}
 \label{fig_outage_vs_PI}
 \end{center}
\end{figure}

\begin{figure}[t!]
\centering
\includegraphics[width= 3.5in, height=2.5in]{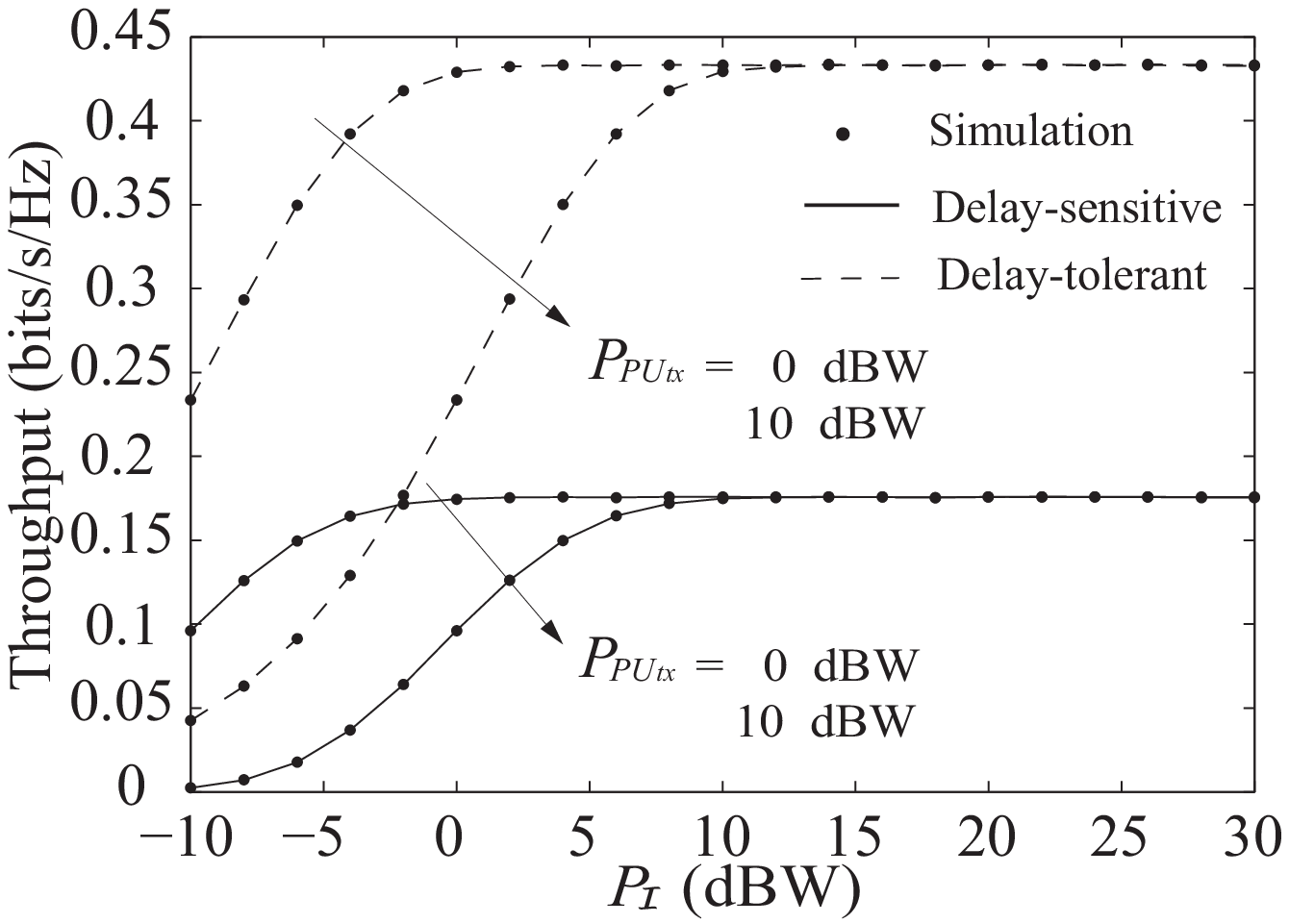}
 \caption{Throughput as a function of $P_{\mathcal{I}}$, with $M=3$, and $\gamma_{th}=0$ dB.}\label{fig_theorughput_VS_PI}
\end{figure}

\begin{figure} [t!]
\begin{center}
\includegraphics[width=3.5in, height=2.5in]{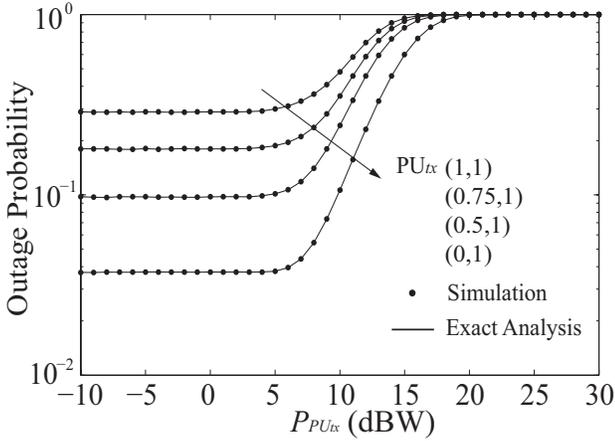}
 \caption{Outage probability as a function of $P_{PU_{tx}}$, with $M=3$, $\gamma_{th}=-10$ dB, and $P_{\mathcal{I}}=10$ dBW.}
 \label{fig_outage_vs_PU}
 \end{center}
\end{figure}

\begin{figure} [t!]
\centering
\includegraphics[width= 3.5in, height=2.5in]{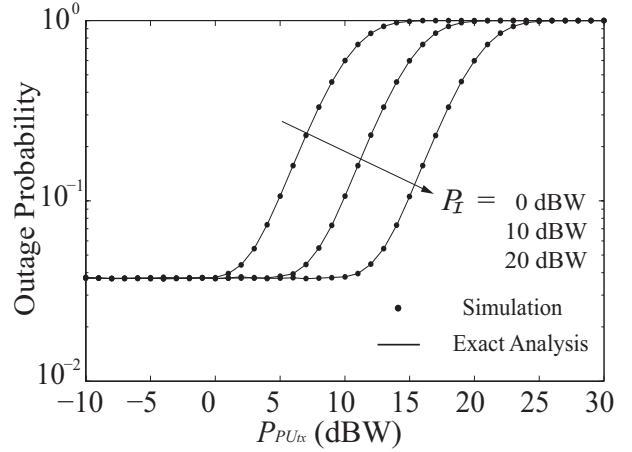}
 \caption{Outage probability as a function of $P_{PU_{tx}}$, with $M=3$ and $\gamma_{th}=-10$ dB.}\label{fig_Pout_vs_Ppu}
\end{figure}

\begin{figure} [t!]
\centering
\includegraphics[width= 3.5in, height=2.5in] {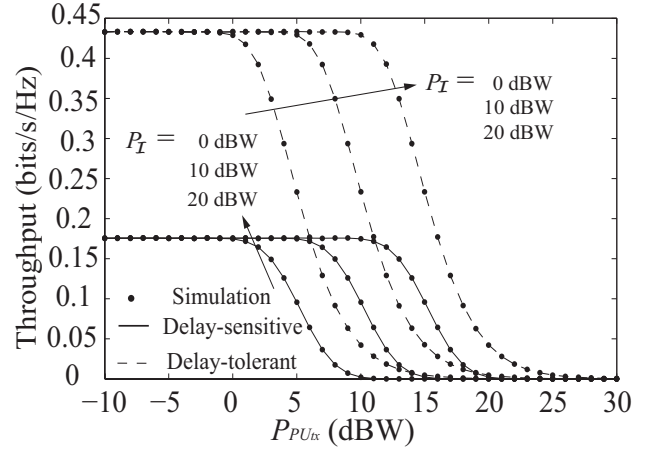}
 \caption{Throughput as a function of $P_{PU_{tx}}$, with $M=3$ and $\gamma_{th}=0$ dB.}\label{fig_throughput_vs_Ppu}
\end{figure}

\begin{figure} [t!]
\centering
\includegraphics[width=3.5in, height=2.5in]{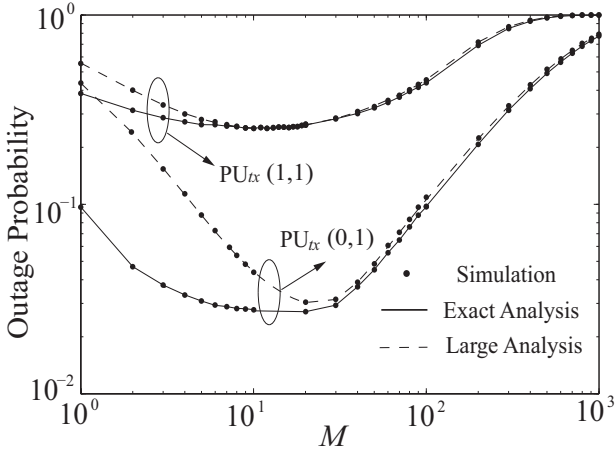}
 \caption{Outage probability as a function of $M$, with $\gamma_{th}=-10$ dB, $P_{\mathcal{I}}=10$~dBW, and $P_{PU_{tx}}=0$~dBW.}\label{fig_outage_vs_MN}
\end{figure}

\begin{figure} [t!]
\centering
\includegraphics[width= 3.5in, height=2.5in]{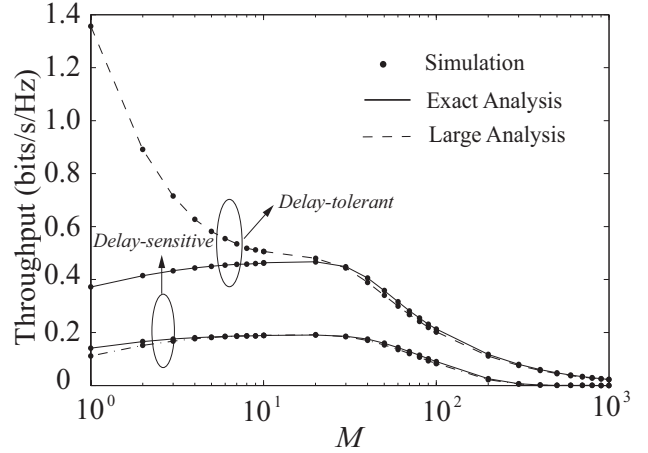}
 \caption{Throughput as a function of $M$, with $\gamma_{th}=0$ dB, $P_{\mathcal{I}}=10$~dBW, and $P_{PU_{tx}}=0$~dBW.}\label{fig_throughput_vs_MN}
\end{figure}

In this section, we present numerical results to examine the outage probability and the throughput for the proposed cognitive relay networks with wireless EH. We show the outage probability and the throughput as functions of $P_{\mathcal{I}}$, $P_{PU_{tx}}$, and the position of $\text{PU}_{tx}$. To gain more insights, we examine the asymptotic behavior of the outage probability and the throughput as the number of PU transceivers goes to infinity. In this work, we assume that all the distances between each nodes are not smaller than one. Without loss of generality, SS, SR, and SD are located at (0,0), (1,0), and (2,0) on the X-Y plane, respectively and the center of PU receivers is located at (2, 1). We specify the location of the center of PU transmitters in the description of each figure. We assume that $\alpha= 0.5$ (Figs. \ref{fig_outage_vs_PI}-\ref{fig_throughput_vs_MN}), $\eta=0.8$, and $M=N$ for all the figures. In each figure, we see excellent agreement between the Monte Carlo simulation points marked as ``$\bullet$'' and the analytical curves, which validates our derivation.

Fig. \ref{fig_outage_vs_PI} shows the outage probability as a function of $P_{\mathcal{I}}$ for various threshold $\gamma_{th}$ values. We assume that the center of $\text{PU}_{tx}$ is located at $(0,1)$. The solid curves, representing the exact analysis, are obtained from \eqref{eq_Pout_simplfied}. Several observations
can be drawn as follows: 1) For a given value of $P_{\mathcal{I}}$, the outage probability decreases as threshold value $\gamma_{th}$ decreases; 2) As the interference power constraint, $P_{\mathcal{I}}$, increases, the outage probability decreases. This is due to the fact that SS and SR can transmit at higher powers without causing interference beyond $P_{\mathcal{I}}$; and 3) The outage probability reaches error floors when $P_{\mathcal{I}}$ is in the high power regime. This is because when $P_{\mathcal{I}}$ goes to infinity, the power constraints at SS and SR are relaxed. The maximum transmit powers at SS and SR are only dependent on the harvested energies.

Fig. \ref{fig_theorughput_VS_PI} shows the throughput in delay-sensitive and delay-tolerant modes as a function of $P_{\mathcal{I}}$ for various transmit power from PU transmitters. The center of $\text{PU}_{tx}$ is located at $(0,1)$. The solid curves, representing the exact analysis of delay-sensitive mode, are obtained from \eqref{eq_Throughput_DL}. The dashed curves, representing the exact analysis of delay-tolerant mode, are obtained from \eqref{eq_Throughput_DT}. The results show that: 1) the throughput increases as $P_{\mathcal{I}}$ increases. This is because SS and SR can transmit at higher transmission power without causing interference beyond $P_{\mathcal{I}}$ as it increases; 2) for different value of $P_{PU_{tx}}$, the throughput can reach the same asymptotic values when $P_{\mathcal{I}}$ is in high power regime. The reason is that as $P_{\mathcal{I}}$ goes to infinity, the power constraints at SS and SR are relaxed. The SIR expressions at SR and SD become independent of the value $P_{PU_{tx}}$; and 3) The throughput results for delay-sensitive mode is lower than those in delay-tolerant. The reason is that in delay-sensitive, the information is transmitted at a fixed rate. If the rate is above the channel rate, the outage occurs and the throughput suffers. While in delay-tolerant transmission mode, the transmission rate is flexible due to the fact that the data at SS can tolerate delays.

Fig. \ref{fig_outage_vs_PU} illustrates the outage probability as a function of $P_{PU_{tx}}$ for various locations of the PU transmitter center.  We observe that: 1) The outage performance degrades with increasing $P_{PU_{tx}}$. This can be explained as follows. On the one hand, as $P_{PU_{tx}}$ increases, SS and SR can harvest more energy from PU transmitters, which can improve the performance of SN. On the other hand, as $P_{PU_{tx}}$ increases, the interference at SR and SD also increases. There exists a tradeoff between the transmit power and performance. The observed simulation results indicate that the detrimental effect of interference from PU transmitters outweighs the benefits of increased energy harvested from the PU transmitters; 2) The outage probability decreases as the center point of $\text{PU}_{tx}$ moves towards SS and away from SR and SD. This is because less interference is imposed on SR and SD while more energy is being harvested at SS, as $\text{PU}_{tx}$ moves towards SS and away from SR and SD. As a result, SS and SR can transmit information successfully at high power levels and hence, the outage probability for a given $P_{PU_{tx}}$ decreases; 3) There exists error floors when $P_{PU_{tx}}$ is in the low power regime, since in this case, the energy that can be harvested at SS and SR is the power constraint which limits the performance; and 4) There exists error ceilings when $P_{PU_{tx}}$ is in the high power regime, since in this case, the fixed interference power constraint $P_{\mathcal{I}}$ limits the transmit power of SS and SR to affect the outage probability. Therefore, if $P_{PU_{tx}}$ goes to infinity, the interference will go to infinity and the outage probability will approach one.

Fig. \ref{fig_Pout_vs_Ppu} shows the outage probability as a function of $P_{PU_{tx}}$ for various $P_{\mathcal{I}}$. We assume $\text{PU}_{tx}$ is located at $(0,1)$. We observe that there exists floors and ceilings of the outage probability similar to those observed in Fig. \ref{fig_outage_vs_PU}. In addition, we observe that the outage probability decreases as $P_{\mathcal{I}}$ increases. This is expected since higher interference power constraint $P_{\mathcal{I}}$ allows higher transmit power at the SS and SR, which results in a lower outage probability. The outage probability curves shift to the right as $P_{\mathcal{I}}$ increases. This behavior can be explained as follows: 1) as $P_{PU_{tx}}$ increases, SS and SR can use higher transmit power to offset the interference from PU transmitters at SR and SD, respectively, and 2) as $P_{\mathcal{I}}$ increases, the transmit power constraints at SS and SR are relaxed, thereby, allowing SS and SR to transmit at higher power levels without exceeding the $P_{\mathcal{I}}$ interference power constraint.

In Fig. \ref{fig_throughput_vs_Ppu}, we plot the throughput for delay-sensitive and delay-tolerant transmission modes as a function of $P_{PU_{tx}}$, where we assume $\text{PU}_{tx}$ is located at $(0,1)$. It is shown that the throughput ceilings and floors exit for $P_{PU_{tx}}$. We observe that the detrimental effect of the interference from $\text{PU}_{tx}$ dominates the throughput performance, such that the throughput decreases as $P_{PU_{tx}}$ increases. We also observe that the throughput curves moves to the right as $P_{\mathcal{I}}$ increases. This is due to the fact that when $P_{\mathcal{I}}$ is relaxed (i.e., increased), SS and SR can transmit at higher power levels to offset the interference from PU transmitters at SR and SD, respectively, as $P_{PU_{tx}}$ increases. Note that similar to our discussion in Fig.~\ref{fig_theorughput_VS_PI}, the throughput in delay-tolerant mode is higher than that in delay-sensitive mode.

Fig. \ref{fig_outage_vs_MN} shows the exact outage probability and asymptotic outage probability as a function of $M$ for various $\text{PU}_{tx}$ positions.  We plot the exact outage probability and asymptotic outage probability curves using \eqref{eq_Pout_simplfied} and \eqref{P_out_large_2}, respectively. The results from the large analysis converge to those from the exact analysis as the number of PU transceivers approaches infinity. There exists an optimal $M$ which can minimize the outage probability. It is shown that the outage probability decreases and then increases as the number of PU transceivers grows large. Since SS and SR can harvest energy from all PU transmitters, it is natural to assume that the opportunity to harvest energy from the PU network increases as the number of PU transmitters increases. In order to improve the system outage probability given by \eqref{eq_Pout_simplfied}, SS and SR can forward information at higher transmit power as the harvested energy improve. However, the harmful impact of the interference imposed on the secondary network from the PN counterbalances the benefits gained from the energy harvesting as the number of PU transmitters increases. Similar to the observation in Fig. \ref{fig_outage_vs_PU}, a lower outage probability can be achieved when $\text{PU}_{tx}$ is far from SD and SR, and close to SS for the each $M$ value. It is also shown that increasing the number of PUs can not guarantee lower outage probability.

In Fig. \ref{fig_throughput_vs_MN}, we plot the throughput for delay-sensitive and delay-tolerant transmission modes as a function of $M$. We assume $P_{\mathcal{I}}=10$~dB, $P_{PU_{tx}}=0$~dB, and $\text{PU}_{tx}$ is located at $(0,1)$. It can be seen that the throughput increases and then decreases as a function of $M$. We observe that the throughput approaches zero as $M$ grows large ($M\sim100$) due to excessive interference power from PU transmitters at SR and SD. The large interference power disables SS and SR from communicating the information to SD. The throughput results from the large analysis converges to those from the exact analysis as $M$ increases.

\begin{figure} [t!]
\centering
\includegraphics[width=3.5in, height=2.5in]{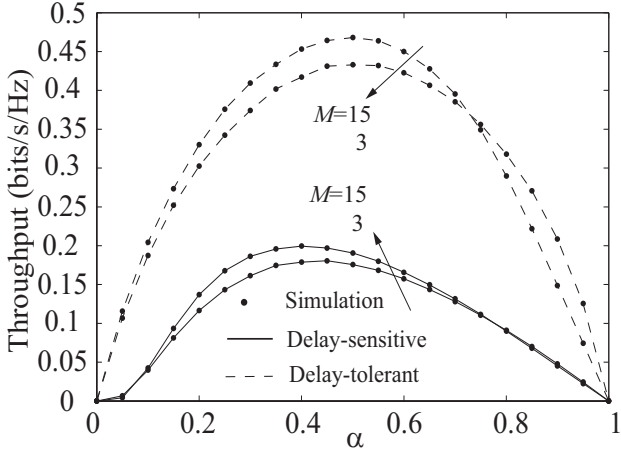}
 \caption{Throughput as a function of $\alpha$, with $\gamma_{th}=0$ dB, $P_{PU_{tx}}=0$~dBW, and $P_{\mathcal{I}}=10$~dBW.}\label{fig_theroughput_vs_alpha}
\end{figure}

In Fig. \ref{fig_theroughput_vs_alpha}, we plot the throughput for delay-sensitive and delay-tolerant transmission modes versus $\alpha$ for various values of $M$. We assume that the PU transmitters are loceted at $(0,1)$.  The results show that the throughout increases and then decreases as a function of $\alpha$. For small values of $\alpha$ ($\sim0$), the SS and SR harvest insufficient energy for a reliable information transmission due to short energy harvesting periods and therefore, the throughput is low. For large values of $\alpha$ ($\sim1$), the SS and SR are unable to transmit reliably due to short information transmitting periods. As $\alpha$ increases from small values to large values, the throughput is influenced by the tradeoff between the energy harvesting and information transmission periods. We observe that the delay-sensitive and delay-tolerant throughput improve when the number of PU transceivers increase from $3$ to $15$. This is due to the fact that the negative effects brought by the interference from PU transmitters on SR and SD offsets the positive effects brought by harvesting energy from PU transmitters on SS and SR. The results also show that the optimal $\alpha$ value is not necessarily the same for the delay-sensitive and delay-tolerant transmission modes.

\textbf{Remark 1:}
The value chosen for $\alpha$ greatly impacts the outage probability and the throughput of the secondary network. As the value of $\alpha$ increases from $0$ to $1$, the duration of EH at SS and SR in each time slot increases and naturally, the remaining duration of the time slot to transmit information decreases. In other words, more energy may be available at SS and SR but less time is available to transmit the information. It is desired to find a value for $\alpha$ which can maximize the throughput. The optimal value of $\alpha$ can be evaluated numerically (and off-line) using search methods for \eqref{eq_Throughput_DL} and \eqref{eq_Throughput_DT}. Note that the closed form expression for the optimal $\alpha$ is not tractable due to the complexity of the throughput expressions.

\textbf{Remark 2:}
It is desired to deploy SS, SR, and SD (with respect to the PN) where the outage probability is minimized and the throughput is maximized. For example, the SN can be deployed so that PU transmitters are located near SS and SR for EH and away from SR and SD to reduce interference power at SR and SD. In order to reduce the interference from SS and SR at the PN, it is desired that SS and SR are far from PU receivers. This can be observed in Fig. \ref{fig_outage_vs_PU}, at the location where the outage probabilities as a function of $P_{PU_{tx}}$ for four $\text{PU}_{tx}$ locations are illustrated. If the system engineer can deploy the SN network at a desired location and $\text{PU}_{tx}$ and $\text{PU}_{rx}$ are immobile, the optimal location of SS, SR, and SD with respect to $\text{PU}_{tx}$ and $\text{PU}_{rx}$ can be obtained. Although a closed-form solution for the optimal location of SS, SR, and SD from \eqref{eq_Pout_simplfied} is intractable, the solution can be obtained offline by numerical searching methods.

\section{Conclusions}\label{sec_concl}
A wireless energy harvesting protocol for an underlay cognitive relay network with multiple primary users (PUs) was proposed. We performed the exact analysis as well as asymptotic analysis as the number of PUs goes to infinity. Expressions for the exact outage probability and the exact throughput for two transmission modes, namely delay-sensitive and delay-tolerant, were derived. For a sufficiently large number of PUs, we derived closed-form asymptotic expressions for the outage probability and the delay-sensitive throughput and an analytical expression for the delay-tolerant throughput. Our results show that the detrimental effect caused by the interference from the multiple PU transmitters at the secondary user (SU) outweighs the benefits brought by harvesting energy when the number of PUs is large. The results also show that as PU transmitters move closer to SS and farther away from SR and SD, more energy is harvested at SS and less interference is imposed on SR and SD so that the outage probability of the SU decreases. In our future work, we will use stochastic geometry to model the locations of PUs and investigate the relationship between density of PUs and performance of SUs.
\appendices
\section{}\label{sec_appendixA}
In this Appendix, we provide a proof of \emph{Theorem 1}. Recall that $\Gamma_{R}$ and $\Gamma_{D}$ are two random variables which are dependent on $Z_{2}$. In order to express \eqref{eq_Pout} in terms of the product of two independent random variables, we condition the term $\Pr\{\Gamma_{R} \geq \gamma_{th}, \Gamma_{D}\geq\gamma_{th}\}$ on $Z_{2}$. Thus,
\begin{eqnarray}\label{P_gamma_RD_appendixA}
\Pr \{ \left. {{\Gamma _R} \ge {\gamma _{th}},{\Gamma _D} \ge {\gamma _{th}}} \right|{Z_2}\} &=& \Pr\{\Gamma_{R} \geq \gamma_{th}|Z_{2}\}\times \nonumber\\
&&\Pr\{\Gamma_{D} \geq \gamma_{th}|Z_{2}\},
\end{eqnarray}

\noindent where $\Pr\{\Gamma_{R} \geq \gamma_{th}|Z_{2}=z_{2}\}$ and $\Pr\{\Gamma_{D} \geq \gamma_{th}|Z_{2}=z_{2}\}$ are two independent probabilities for a given $z_2$. We derive and simplify the expressions for $\Pr\{\Gamma_{R}\geq \gamma_{th}|Z_{2}=z_{2}\}$ and $\Pr\{\Gamma_{D}\geq \gamma_{th}|Z_{2}=z_{2}\}$. The term $\Pr\{\Gamma_{R}\geq \gamma_{th}|Z_{2}=z_{2}\}$ can be obtained as
\begin{eqnarray}
&&\Pr\{\Gamma_{R}\geq\gamma_{th}|Z_{2}\}\nonumber\\
\hspace{+.5cm}&=&\Pr\bigg\{\min\big( \rho Z_{1}, \frac{P_{\mathcal{I}}}{Y_{1}}\big)\frac{X_{1}}{Z_{2}}\geq\gamma_{th}\bigg\}\label{eq_CDF_gamma_R_s1_der_l1}\\
\hspace{+.5cm}&=&\underbrace{\Pr\bigg\{X_{1}\geq \frac{Z_{2}\gamma_{th}}{Z_{1}\rho}, Y_{1}\leq \frac{P_{\mathcal{I}}}{Z_{1}\rho}\bigg\}}_{\mathcal{J}_{R,I}} \nonumber\\
\hspace{+.5cm}&+& \underbrace{\Pr\bigg\{X_{1}\geq \frac{Y_{1}Z_{2}\gamma_{th} }{P_{\mathcal{I}}}, \frac{P_{\mathcal{I}}}{Y_{1}\rho}\leq Z_{1}\bigg\}}_{\mathcal{J}_{R,II}},
\label{eq_CDF_gamma_R_s1_der}
\end{eqnarray}
\noindent where \eqref{eq_CDF_gamma_R_s1_der_l1} presents the probability that SIR is greater than $\gamma_{th}$ based on the three important power constraints. We present \eqref{eq_CDF_gamma_R_s1_der_l1} in terms of the addition of $\mathcal{J}_{R,I}$ and $\mathcal{J}_{R,II}$ based on two possibilities: $\rho Z_{1}<\frac{P_{\mathcal{I}}}{Y_{1}}$ and $\rho Z_{1}>\frac{P_{\mathcal{I}}}{Y_{1}}$.
Conditioning $\mathcal{J}_{R,I}$ in (\ref{eq_CDF_gamma_R_s1_der}) on $Z_{1}$ and taking the expected value of the results over the distribution of $Z_{1}$, we have
\begin{eqnarray}
\mathcal{J}_{R,I}&=&\int_{0}^{\infty} e^{-\frac{Z_{2}\gamma_{th}}{z_{1}\rho \lambda_{1}}}(1-e^{-\frac{P_{\mathcal{I}}}{z_{1}\rho \omega_1}})^{M}f_{Z_{1}}(z_{1})dz_{1}\nonumber\\
&=& \int_{0}^{\infty} e^{-\frac{Z_{2}\gamma_{th}}{z_{1}\rho \lambda_{1}}}(1-e^{-\frac{P_{\mathcal{I}}}{z_{1}\rho \omega_1}})^{M}\nonumber\\
&& \;\;\;\;\;\;\times\frac{z_{1}^{N-1}e^{-\frac{z_{1}}{P_{PU_{tx}}\nu_{1}}}}{\Gamma(N)(P_{PU_{tx}}\nu_{1})^{N}}dz_{1}.
\label{eq_J_R_I}
\end{eqnarray}

Note that by conditioning $\mathcal{J}_{R,I}$ on $Z_{1}$, the two terms in $\mathcal{J}_{R,I}$ become independent with respect to one another and therefore, $\mathcal{J}_{R,I}$ can be presented as the product of the two terms.
Similarly, by conditioning $\mathcal{J}_{R,II}$ in (\ref{eq_CDF_gamma_R_s1_der}) on $Y_{1}$ and by taking the expected value of the result over the distribution of $Y_{1}$, we have
\begin{eqnarray}
\mathcal{J}_{R,II}&=& \int_{0}^{\infty}
e^{-\frac{y_{1}Z_{2}\gamma_{th} }{P_{\mathcal{I}}\lambda_{1}}} \big(1-F_{Z_{1}}(\frac{P_{\mathcal{I}}}{y_{1}\rho })\big) f_{Y_{1}}(y_{1})dy_{1}\nonumber\\
&=&  \int_{0}^{\infty}
e^{-\frac{y_{1}Z_{2}\gamma_{th}}{P_{\mathcal{I}}\lambda_{1}}} \big(1-\frac{\Gamma(N,\frac{P_{\mathcal{I}}}{P_{PU_{tx}}\nu_{1}y_{1}\rho })}{\Gamma(N)}\big)\nonumber\\ && \;\;\;\;\;\;\times\frac{M}{\omega_{1}} \sum_{k}^{M-1} {M-1 \choose k} (-1)^{k}e^{-(\frac{k+1}{\omega_{1}})y_{1}}dy_{1}.\nonumber\\
\label{eq_J_R_2}
\end{eqnarray}
In the same manner, the term $\Pr\{\Gamma_{D}\geq \gamma_{th}|Z_{2}\}$ in \eqref{eq_Pout_simplfied} can be obtained as
\begin{eqnarray}
&&\Pr\{\Gamma_{D}\geq\gamma_{th}|Z_{2}\}\nonumber\\
&=& \Pr\bigg\{\min\big( \rho Z_{2}, \frac{P_{\mathcal{I}}}{Y_{2}}\big)\frac{X_{2}}{Z_{3}}\geq\gamma_{th}\bigg\}\nonumber\\
&=&\underbrace{\Pr\bigg\{X_{2}\geq \frac{\gamma_{th} Z_{3}}{Z_{2}\rho}, Y_{2}\leq \frac{P_{\mathcal{I}}}{Z_{2}\rho}\bigg\}}_{\mathcal{J}_{D,I}} \nonumber\\
&+& \underbrace{\Pr\bigg\{X_{2}\geq \frac{\gamma_{th} Y_{2}Z_{3}}{P_{\mathcal{I}}}, Y_{2}\geq \frac{P_{\mathcal{I}}}{Z_{2}\rho}\bigg\}}_{\mathcal{J}_{D,II}}.
\label{eq_CDF_gamma_D_s1_der}
\end{eqnarray}
Since $\mathcal{J}_{D,I}$ is conditioned on $Z_{2}$, the joint probability can be written as the product of two marginal probabilities, i.e.,
\begin{eqnarray}
\mathcal{J}_{D,I}&=&\underbrace{\Pr\{X_{2}\geq \frac{\gamma_{th}Z_{3}}{\rho Z_{2}}\}}_{\mathcal{I}_{1}}\underbrace{\Pr\{Y_{2}\leq \frac{P_{\mathcal{I}}}{\rho Z_{2}}\}}_{\mathcal{I}_2}, \nonumber\\
&=& \frac{{{{\left( {1 - {e^{ - \frac{{{P_{\cal I}}}}{{{Z_2}\rho {\omega _2}}}}}} \right)}^M}}}{{{{\left( {1 + \frac{{{P_{P{U_{tx}}}}{\nu _3}{\gamma _{th}}}}{{\rho {Z_2}{\lambda _2}}}} \right)}^N}}},
\label{eq_J_D_I}
\end{eqnarray}
\noindent where $\mathcal{I}_{1}$ is obtained by applying \cite[Eq 3.326.2]{TablOfIntegrals}
\begin{eqnarray}
\mathcal{I}_{1}&=&\int_{0}^{\infty} e^{-\frac{\gamma_{th} z_{3}}{\rho Z_{2}\lambda_{2}}}f_{Z_{3}}(z_{3})dz_{3}\nonumber\\
&=& {\left( {1 + \frac{{{P_{P{U_{tx}}}}{\nu _3}{\gamma _{th}}}}{{\rho {Z_2}{\lambda _2}}}} \right)^{ - N}},
\label{eq_I_1}
\end{eqnarray}
and
\begin{equation}
\mathcal{I}_{2}=(1-e^{-\frac{P_{\mathcal{I}}}{Z_{2}\rho\omega_{2}}})^{M}.
\label{eq_I_2}
\end{equation}
We condition $\mathcal{J}_{D,II}$ on $Z_{3}$
\begin{align}\label{eq_J_D_II_Z3}
&\mathcal{J}_{D,II}|Z_{3}=\Pr\{X_{2}\geq \frac{Y_{2}Z_{3}\gamma_{th}}{P_{\mathcal{I}}},Y_{2}\geq\frac{P_{\mathcal{I}}}{\rho Z_{2}}\}\nonumber\\
&=\int_{\frac{P_{\mathcal{I}}}{\rho Z_{2}}}^{\infty} e^{-\frac{y_{2}Z_{3}\gamma_{th}}{P_{\mathcal{I}}\lambda_{2}}} \frac{M}{\omega_{2}}\sum_{k}^{M-1} {M-1 \choose k} (-1)^{k}e^{-(\frac{k+1}{\omega_{2}})y_{2}} dy_{2}
\end{align}

Averaging $\mathcal{J}_{D,II}$ over the PDF of $Z_{3}$ we have,
\begin{eqnarray}
\mathcal{J}_{D,II}&=& \int_{0}^{\infty} \mathcal{J}_{D,II}|Z_{3}f_{Z_{3}}(z_{3})dz_{3}\nonumber\\
&=&\int_{\frac{P_{\mathcal{I}}}{\rho Z_{2}}}^{\infty} \frac{M}{\omega_{2}} \sum_{k}^{M-1} {M-1 \choose k}(-1)^{k}e^{-(\frac{k+1}{\omega_{2}})y_{2}} \nonumber\\  &&\;\;\times{\frac{1}{{{{\left( {1 + {y_2}\frac{{{\gamma _{th}}{P_{P{U_{tx}}}}{\nu _3}}}{{{P_{\cal I}}{\lambda _2}}}} \right)}^N}}}}dy_{2}.
\label{eq_J_D_II}
\end{eqnarray}
The outage probability can be evaluated using the results obtained from \eqref{eq_J_R_I}, \eqref{eq_J_R_2}, \eqref{eq_J_D_I}, and \eqref{eq_J_D_II} in \eqref{eq_Pout_simplfied}.

\bibliographystyle{IEEEtran}
\balance
\bibliography{harvest_biblio}
\vspace{-1cm}

\end{document}